\documentclass{article}
\usepackage{graphics}
\usepackage{bm}
\usepackage{graphicx}
\usepackage{amsmath}
\usepackage{amssymb}
\usepackage{amstext}
\usepackage{amscd}
\usepackage{amsfonts}
\usepackage{hyperref}
\usepackage{textcomp}
\usepackage{color}
\DeclareMathAlphabet{\EuFrak}{U}{euf}{m}{n}
\DeclareMathAlphabet{\EuScript}{U}{eus}{m}{n}

\newcommand{\be}{\begin{equation}}
\newcommand{\ee}{\end{equation}}
\newcommand{\ben}{\begin{eqnarray}}
\newcommand{\een}{\end{eqnarray}}

\date{\today}

\begin{document}

\title{{\bf Finite Tsallis Gravitational Partition Function for a System of Galaxies}}
\author{{M. Hameeda$^{1,2,a}$, B. Pourhassan$^{3,4,b}$, M.C.Rocca$^{5,6,7,c,d}$}\\
{\texttt{ \rm{M. Faizal$^{8,9,e}$}}}\\
\small{$^1$ Department of Physics, S.P. College, Srinagar, Kashmir, 190001 India}\\
\small{$^2$ Inter University Centre for Astronomy and Astrophysics , Pune India}\\
\small{$^3$ School of Physics, Damghan University,}\\
\small{ P. O. Box 3671641167, Damghan, Iran}\\
\small{$^4$ Canadian Quantum Research Center 204-3002 32 Ave Vernon, }\\
\small{BC V1T 2L7 Canada}\\
\small{$^5$ Departamento de F\'{\i}sica,
Universidad Nacional de La Plata,}\\
\small{$^6$ Departamento de Matem\'{a}tica,
Universidad Nacional de La Plata,}\\
\small{$^7$ Consejo Nacional de Investigaciones Cient\'{\i}ficas
y Tecnol\'{o}gicas}\\
\small{(IFLP-CCT-CONICET)-C. C. 727, 1900 La Plata -
Argentina}\\
\small{$^8$Irving K. Barber School of Arts and Sciences, University of British Columbia,}\\
\small{Kelowna, British Columbia, V1V 1V7, Canada}\\
\small{$^9$Department of Physics and Astronomy, University of Lethbridge,}\\
\small{ Lethbridge, Alberta, T1K 3M4, Canada}\\
\small{\texttt{\rm{$^{a}$hme123eda@gmail.com, $^{b}$b.pourhassan@du.ac.ir,}}}\\
\small{\texttt{\rm{$^{c}$rocca@fisica.unlp.edu.ar,$^{d}$mariocarlosrocca@gmail.com, }}}\\
\small{\texttt{\rm{$^{e}$mirfaizalmir@googlemail.com}}}}

\maketitle

\begin{abstract}
In this paper, we  will study the large scale structure formation using the gravitational partition function. We will assertively argue that the system of gravitating galaxies can be analyzed using the Tsallis statistical mechanics.  The divergences in the Tsallis gravitational partition function can be removed using the mathematical riches of the generalization of the dimensional regularization (GDR). The finite gravitational partition function thus obtained will be used to evaluate the thermodynamics of the system of galaxies and thus, to understand the clustering of galaxies in the universe. The correlation function which is believed to contain the information of clustering of galaxies will also be discussed in this formalism.
\end{abstract}

\maketitle

\newpage

\tableofcontents

\newpage

\renewcommand{\theequation}{\arabic{section}.\arabic{equation}}

\section{Introduction}
It is known that the large scale structure of our universe is formed due to clustering of galaxies
 \cite{PEEBLES,   clus}.
 As the size of individual galaxies is small as compared to the distance between them, we can approximate individual galaxies as point particles, and use the standard techniques of statistical mechanics
to construct a  gravitational partition function \cite{sas, sas2}.
 However, by  approximating the galaxies as point particles (and neglecting their extended structure), we end up with divergences.
These divergences have been removed by adding a  softening parameter by hand \cite{ahm10}. This softening parameter is expected to occur due to  the extended structure of galaxies. So, it has been suggested that if a cut-off is introduced than the gravitational partition function would not diverge. However, this has not been explicitly demonstrated.
So, in this letter we will use the generalization of the dimensional regularization (GDR) \cite{jpco,xz16,la12, la14, la16, la18} to obtain a finite or divergence free gravitational partition function.
This method generalize the dimensional regularization of Bollini and Giambiagi \cite{xz12,xz14,xz15,xz18,di12,di14}
It may be noted that the entropy obtained from the gravitation partition function has been used to analyze the  clustering of galaxies.  This has been done by relating the entropy of the system of galaxies to the clustering parameter, and this can in turn be related to the observations  \cite{sas84, ahm02, 10, 20, ahm06}.

It may be considered that the accelerated expansion of the universe \cite{1ab, 6ab} can have important consequences for the formation of large scale structure in the universe, and so the modification of gravitational partition function from a cosmological constant has also been studied  \cite{1b}. As it is possible to consider models with time dependent cosmological constant, the modification of the gravitational partition function from a time dependent cosmological constant has also been studied  \cite{1}. Here again the modified gravitational partition function is used to obtain the entropy of the system, which is then related to the clustering parameter, and hence the effects of the cosmological constant on the clustering are studied.

It is known that a large scale modification of gravitational potential occur by adding higher powers of scalar curvature to general relativity  \cite{gravity}.

Such a modification of the gravitational potential can cause a modification of gravitation partition function \cite{gravity}. This modified gravitational partition function has been used to obtain the entropy of a system of galaxies in such a modified theory of gravity, and this corrected entropy  has been used to analyze the clustering of such a system of galaxies \cite{2}.  Such modification to gravitational potential has  been constrained from the observational data  \cite{2az}. The entropy  of a system of galaxies interacting through a modified Newtonian dynamics (MOND) modified gravitational partition function has also been studied   \cite{3}.   It was demonstrated  that the   MOND corrections modified  the entropy of this system, and the  corrected entropy changed  the clustering parameter for this  system.   The gravitational partition function in MOG has also been studied, and used  to analyze the clustering of galaxies  using the corrected entropy of the system.
The corrected entropy of  a system of galaxies  in brane world models has been obtained from    gravitational partition function \cite{Hameeda2}. This has been obtained using the modification to the gravitational potential from  super-light brane world perturbative modes.

A detailed review of f(R)-cosmological model presented in \cite{capo} is interesting not only for being exactly integrable but for being able to explain the decelerated phase and accelerated phase under the same standard. The two phases correspond to dust matter phase and dark energy phase respectively. The f(R) model \cite{capo} is able to fulfill many observational tests without consideration of any dark matter \cite{capo2}. The physical non-equivalence of Jordan frame and Einstein frame is explained considering f(R) model which permits one to compare analytically the two frames showing the physical differences. However, the partition function as obtained in this paper being non-relativistic is independent of frame. The same may be the case with the thermodynamics obtained from the corresponding partition function. The frame dependence may be seen in the relativistic situations.
\normalcolor
The gravitational partition function has been used to analyze the gravitational phase transition  \cite{2b}. In fact, it has been demonstrated that such  a phase transition can be analyzed using complex fugacity for such a system of galaxies (using the Yang-Lee theory)  \cite{4b}.  It has also been possible to obtain
cosmic energy equation  from such a gravitational partition function   \cite{Ahmad}. To remove the divergences in  the cosmic energy equation  from extended structure of galaxies, it is possible to modify  the cosmic energy equation using a softening parameter, which acts as cut-off in the gravitational partition function  \cite{ce,  Hameeda}.

Now almost all the work on clustering of galaxies has been done using the entropy obtained from the gravitational partition function. However, this  Boltzmann-Gibbs entropy is based on the  extensive property of the system, and it is possible for extensive property to get violated for those complex system which  violate the   probabilistic   quasi-independence \cite{t1}. This violation occurs due to the fact that there is breakdown of ergodicity for such complex system \cite{t2}. Such a  violation of extensive property occurs for a self gravitating systems \cite{t4}. This is because for such a self gravitational system, the  total energy increase  much faster than the particle number, and this can make the partition function   complex. So, for such system, it is possible to use  Tsallis statistical mechanics,  it is a generalization of the Boltzmann-Gibbs statistical mechanics, which does not require  probabilistic   quasi-independence of the system  \cite{t, t8}. As   Tsallis statistical mechanics can be used to study system which violate the  extensive property of the system, it can be used to analyze a system of self interacting gravitational particles \cite{t6}. Now the galaxies can be approximated as self interacting point particles, so it is possible to use Tsallis statistical mechanics to analyze them. The divergences in the partition function are removed using GDR. Thus we see Tsallis statistical mechanics supplemented by GDR is the wonderful treatment to study the gravitational clustering of galaxies. So, in this paper, we will analyze the formation  of large scale structure of the universe using Tsallis statistical mechanics.

\section{Tsallis Partition Function of Galaxies }

In this section, we will apply   Tsallis statistical mechanics to a system of galaxies.
The partition function for a system of galaxies     has been analyzed using the  Boltzmann-Gibbs statistics \cite{sas, sas2}.
However,  such a self gravitational system can violate   probabilistic   quasi-independence \cite{t6}, so the usual Boltzmann-Gibbs cannot be applied to it, and we need to analyze this system using the
  Tsallis statistical mechanics  \cite{t8}.
 It may be noted that we will observe that this  Tsallis partition function diverges, for a system of galaxies  interacting through a gravitational interaction, and so we will use GDR to obtain a finite expression for it. The Tsallis $q$-exponential,is  defined as
\begin{equation}
\label{eq2.1}
e_q(x)=[1+(q-1)x]_+^{\frac {1} {q-1}}
\end{equation}
Or equivalently
\begin{equation}
\label{ep2.2}
e_q(x)=
\begin{cases}
1+(q-1)x]^{\frac {1} {q-1}}\;\;\;;\;\;\;1+(q-1)x>0\\
0\;\;\;;\;\;\;1+(q-1)x<0
\end{cases}
\end{equation}
Let's consider the distribution
$\frac {1} {r}\equiv PV\frac {1}{r}$.
Then $\frac {1} {r}\mid_{r=0}=0$.
We will express the Tsallis gravitational partition function in
$\nu$ dimensions  as (with $q>1$):
\begin{equation}
\label{eq2.3}
{\cal Z}_\nu=\int\limits_{-\infty}^{\infty}d^\nu x\int\limits_{-\infty}^{\infty}d^\nu p
\left[1+(q-1)\beta\left(\frac {N(N-1)Gm^2} {2r}-\frac {Np^2} {2m}\right)\right]_+^{\frac {1} {q-1}}
\end{equation}
Here we have assumed that all the $N$ galaxies have equal masses $m$. We have also assumed that the potential energy between these galaxies can be denoted by  $\phi$,  their momenta by $p_{i}$,  and average temperature of the system by $T$.
Now we  can   express this Tsallis partition function as:
\begin{equation}
\label{eq2.4}
{\cal Z}_\nu=\left[\frac {2\pi^{\frac {\nu} {2}}} {\Gamma\left(\frac {\nu} {2}\right)}\right]^2
\int\limits_0^{\infty}r^{\nu-1}dr\int\limits_0^{\infty}p^{\nu-1} dp
\left[1+(q-1)\beta\left(\frac {N(N-1)Gm^2} {2r}-\frac {Np^2} {2m}\right)\right]_+^{\frac {1} {q-1}}
\end{equation}
It may be noted that
\begin{equation}
\label{eq2.5}
1+(q-1)\beta\left(\frac {N(N-1)Gm^2} {2r}-\frac {Np^2} {2m}\right)>0,
\end{equation}
and as a consequence
\begin{equation}
\label{eq2.6}
p<\sqrt{\frac {2m} {(q-1)\beta N}+\frac {2m^3(N-1)G} {2r}}=P_0
\end{equation}
Thus, we can express Tsallis gravitation partition function for a system of galaxies as
\begin{eqnarray}
\label{eq2.7}
{\cal Z}_\nu &=&
 \frac {1} {2}\left[\frac {2\pi^{\frac {\nu} {2}}} {\Gamma\left(\frac {\nu} {2}\right)}\right]^2
 \left[\frac {m} {2\beta(q-1)N}\right]^{\frac {\nu} {2}}
[(q-1)\beta N(N-1)Gm^2]^\nu\nonumber\\
&\times& B\left(1+\frac {1} {q-1},\frac {\nu} {2}\right)
B\left(\frac {\nu} {2}-\frac {1} {q-1},-\nu\right)
\end{eqnarray}
It may be noted that as  we  set $q= {4} /{3}$, and $\nu=3$, this partition function diverges.   This divergences has to be removed before this partition function can be used to analyze clustering of galaxies.
The choice of $q={4}/{3} $ is not whimsical. $q={4}/{3}$ is the value of $q $ for which, with the Tsallis statistics,
it was possible to prove the validity of the hypothesis of the emergent gravity of Verlinde \cite{ver}, in the non-relativistic case
\cite {prg}.

\setcounter{equation}{0}

\section{Generalization of the Dimensional Regularization}

It may be noted that as the Tsallis partition function diverges, we need to obtain a finite expression for it. So, to
 write an explicit expression  for the Tsallis gravitational partition function for galaxies in three dimensions, we will use GDR to have a finite result for it \cite{jpco,xz16,la12,la14,la16,la18}. It is a generalization of the usual dimensional regularization, based  on Ultradistributions theory of Sebastiao e Silva, also known as Ultrahyperfunctions.
\cite{jss,hasumi,tp8}.As  the divergences  occur, due to  the  products of distributions with
coincident point singularities, it is possible to use Ultrahyperfunctions  to eliminate them.
These divergences are manifested, for example, in the integrals that describe the systems considered.
As a consequence we can apply this method to obtain finite results for divergent integrals.
We must clarify that this is not a regularization method. It is an exact method based on the theory of Ultrahyperfunctions.
So, now we  define a  function  $f$ as
\begin{equation}
\label{eq3.1}
f(\nu)=-\frac {12\pi^\nu} {\Gamma\left(\frac {\nu} {2}\right)}
\frac {\Gamma\left(\frac {\nu} {2}-3\right)} {\nu(\nu-1)(\nu-2)}\sin\pi\left(\frac {\nu} {2}+4\right)
\left(\frac {mN\beta} {2}\right)^{\frac {\nu} {2}}[m^2G(N-1)]^\nu,
\end{equation}
With this function we can write
\begin{equation}
\label{eq3.2}
{\cal Z}_\nu=f(\nu)\Gamma(3-\nu)
\end{equation}
So, we can express $f(3)$ and $f'(3)$ as
\begin{eqnarray}
\label{eq3.3}
f(3)=-\frac {16\pi^2} {\sqrt{3}}\left(\frac {mN\beta} {2}\right)^{\frac {3} {2}}
[m^2G(N-1)]^3\\
f^{'}(3)=\frac {f(3)} {2}\left\{\ln\{mN\beta[m^2G(N-1)]^2\}-\ln 6-\frac {95} {9}\right\}
\end{eqnarray}
We can write Laurent's expansion for $f$ around $\nu=3$ in the form
\begin{equation}
\label{eq3.4}
f(\nu)=f(3)+f^{'}(3)(\nu-3)+\sum\limits_{k=2}^\infty b_k(\nu-3)^k\\
\end{equation}
Laurent's development of the gamma function is \cite{tp13}:
\begin{equation}
\label{eq3.5}
\Gamma(3-\nu)=\frac {1} {3-\nu}-C+\sum\limits_{k=1}^\infty c_k\left(\nu-3\right)^k
\end{equation}
(where $C$ is the Euler-Mascheroni constant).
Multiplying the Laurent expansions of $f$ and the gamma function, we obtain the expansion of ${\cal Z}$
\begin{equation}
\label{eq3.6}
{\cal Z}_\nu=\frac {f(3)} {3-\nu}-f(3)C-f^{'}(3)+
\sum\limits_{k=1}^{\infty}a_k\left(\nu-3\right)^k.
\end{equation}
The value of the integral is the independent term of the powers of $\nu-3$.
Thus, the Tsallis gravitational partition function for galaxies in three dimensions can be expressed as
\begin{equation}
\label{eq3.7}
{\cal Z}=-f(3)C-f^{'}(3)
\end{equation}
So, we can obtain the finite  Tsallis gravitational partition function for galaxies as
\begin{eqnarray}
\label{eq3.8}
{\cal Z}=\frac {8\pi^2}{3}\left(\frac {m\beta N} {2}\right)^{\frac{3}{2}}
m^2G(N-1)^3 \left\{\ln\{m^{5}N\beta[\pi G(N-1)]^2\}+2C-\ln 6-\frac {95} {9}\right\}
\end{eqnarray}
Now we will write this  Tsallis gravitational partition function for galaxies  as
\begin{equation}
\label{eq3.9}
{\cal Z}=\frac {8\pi^2} {3}\alpha_1(N,\beta)\alpha_2(N,\beta)
\end{equation}
where we have defined $\alpha_1(N,\beta)$ and $\alpha_2(N,\beta)$ as
\begin{eqnarray}
\label{eq3.10}
\alpha_1(N,\beta)&=&\left(\frac {m\beta N} {2}\right)^{\frac {3} {2}}
m^2G(N-1)^3
\\
\alpha_2(N,\beta)&=&\left\{\ln\{mN\beta[\pi m^2G(N-1)]^2\}+2C-\ln 6-\frac {95} {9}\right\}
\end{eqnarray}
It may be noted that the divergences have been removed from this Tsallis gravitational partition function for galaxies. Now this can be used to study various aspects of this system.

\setcounter{equation}{0}

\section{Thermodynamics}

\begin{figure}[h!]
 \begin{center}$
 \begin{array}{cccc}
\includegraphics[width=40 mm]{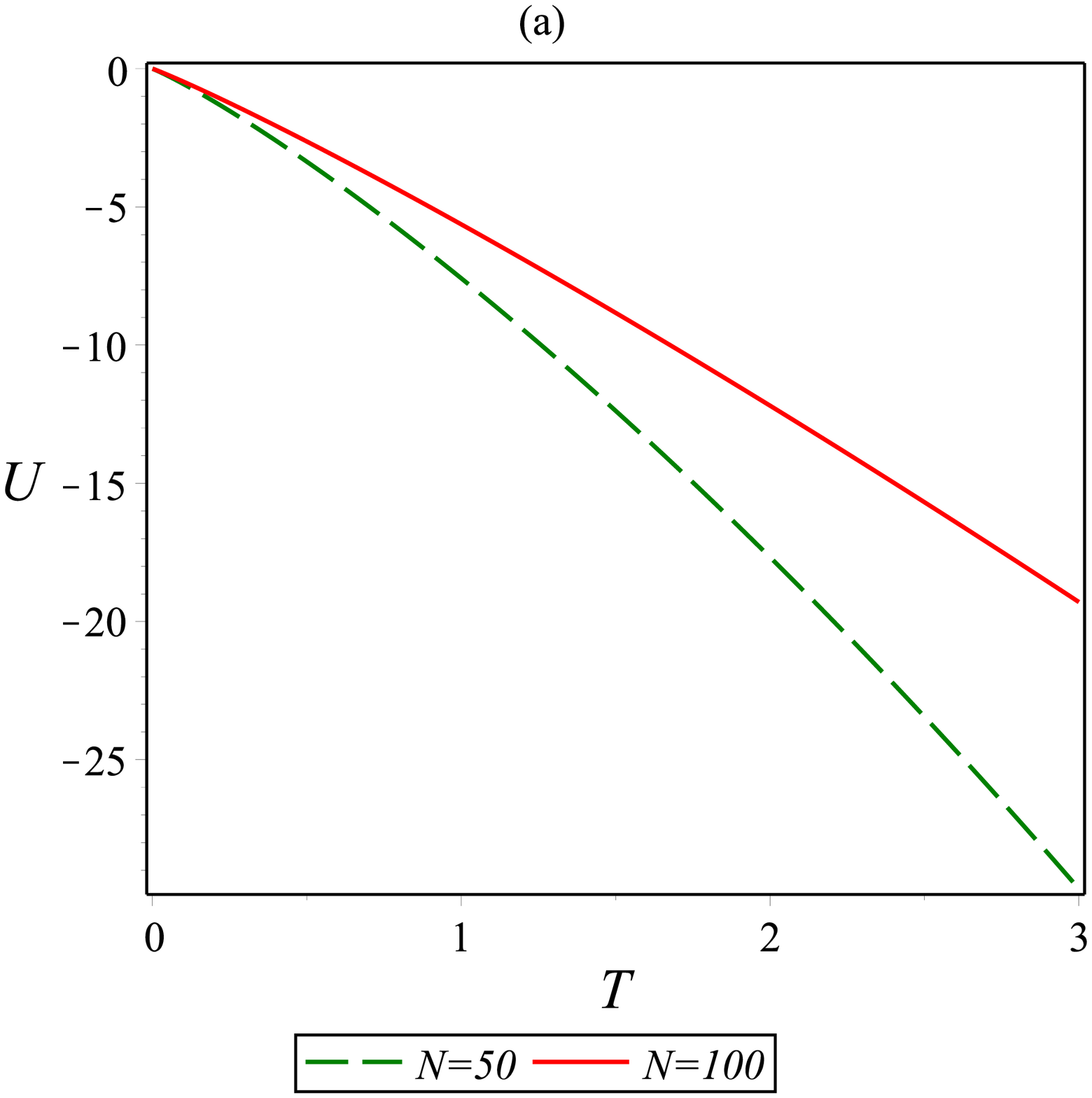}\includegraphics[width=40 mm]{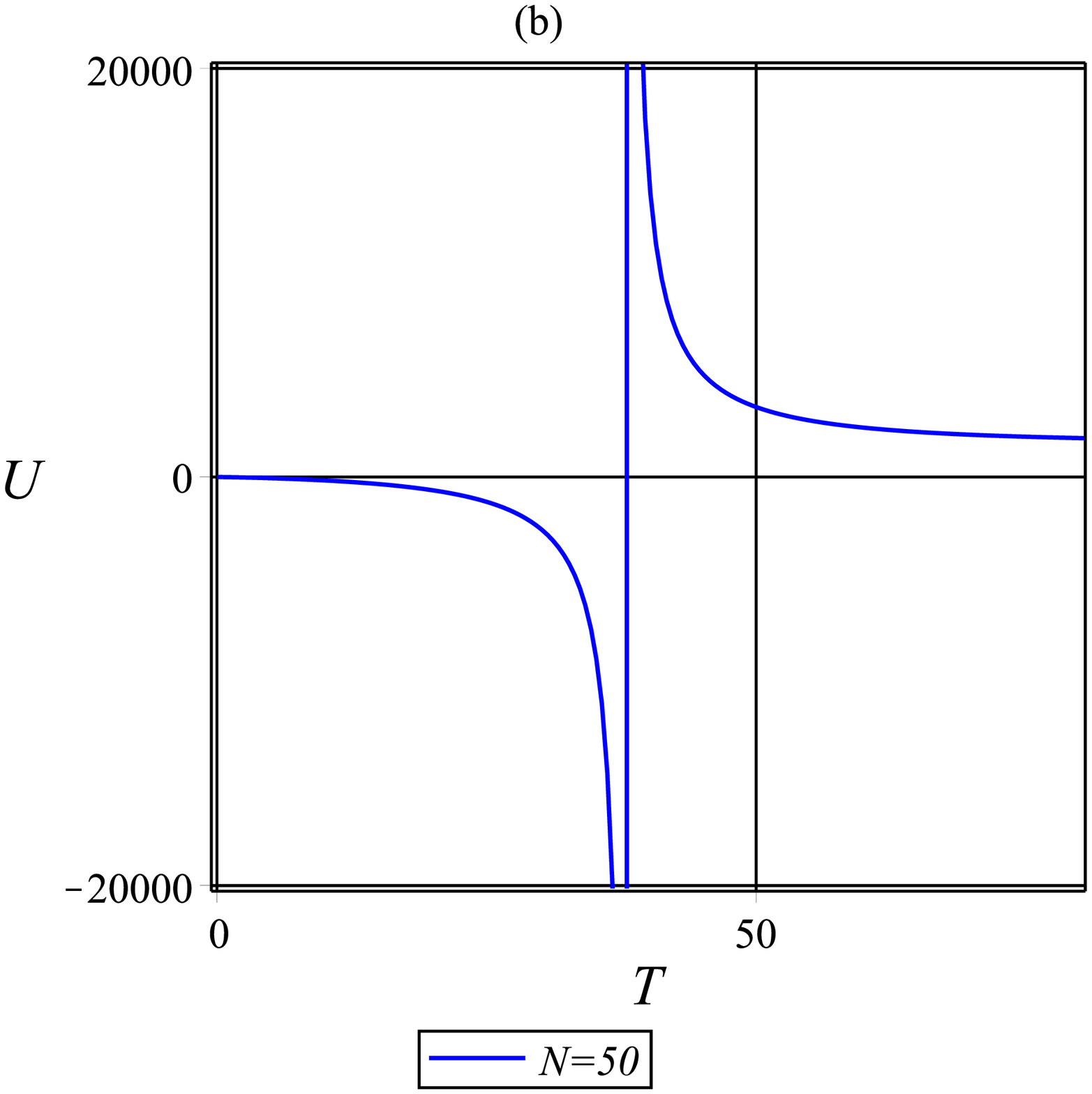}\includegraphics[width=40 mm]{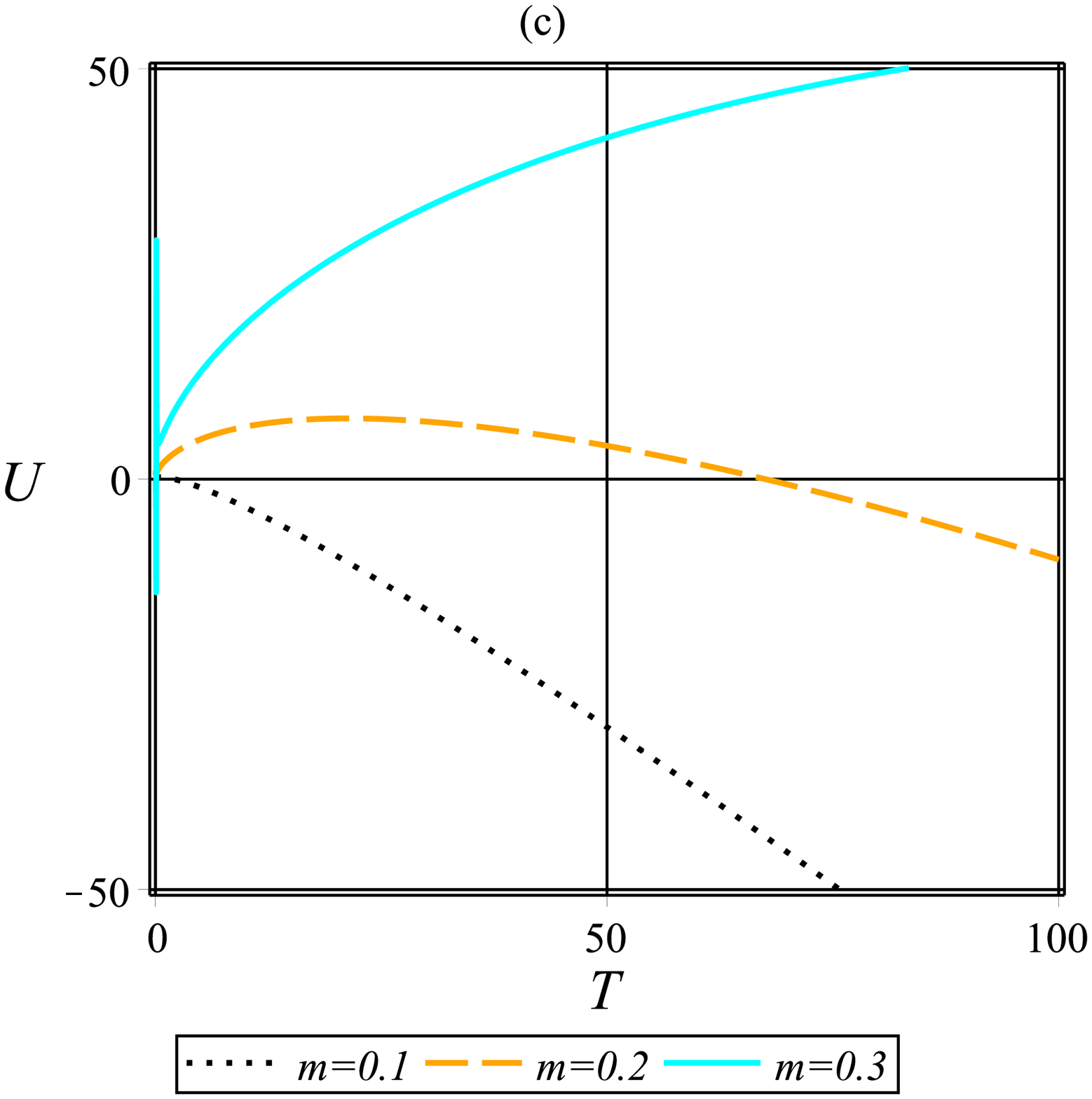}
 \end{array}$
 \end{center}
\caption{Typical behavior of the internal energy in terms of temperature for $G=1$. We set $m=1$ in (a) and (b) plots, while vary $m$ in (c) plot.}
 \label{figu}
\end{figure}

The thermodynamics of a system of galaxies can be studied using the  Tsallis gravitational partition function. This can be done by first using the partition function for obtaining the internal energy of this system, and then using the internal energy to obtain various other thermodynamic quantities for this system. Thus, using this Tsallis partition function, the internal energy for a system of galaxies can be written as
\begin{eqnarray}
{<{\cal U}>_\nu=\frac{1}{\cal Z}}\left[\frac {2\pi^{\frac {\nu} {2}}} {\Gamma\left(\frac {\nu} {2}\right)}\right]^2
\int\limits_0^{\infty}r^{\nu-1}dr
\int\limits_0^{P_0} p^{\nu-1} dp
\left(\frac {N(N-1)Gm^2} {2r}-\frac {Np^2} {2m}\right)
\nonumber \\ \times
\left[1+(q-1)\beta\left(\frac {N(N-1)Gm^2} {2r}-\frac {Np^2} {2m}\right)\right]^{\frac {1} {q-1}}
\end{eqnarray}
So,   we can express internal energy of a system of galaxies as
\begin{eqnarray}
<{\cal U}>&=&-\frac {6\pi^2} {\beta{\cal Z}}\left(\frac {m\beta N} {2}\right)^{\frac {3} {2}}
m^{2}G(N-1)^3\nonumber\\
&\times&\left\{\ln\{m N\beta[\pi m^2G(N-1)]^2\}+2C-\ln 6-\frac {29} {15}\right\}
\end{eqnarray}
After simplification, we can write this expression as following,
\begin{equation}
<{\cal U}>=-\frac{9}{4\beta}\left(1+\frac{388}{45\alpha_2(N,\beta)}\right)
\end{equation}
Thus, the internal energy depends on the value of $\alpha_2(N,\beta)$. As the value of $\alpha_2(N,\beta)$ increases, the value of internal energy decreases. Now it is known that the internal energy decreases as the galaxies cluster \cite{sas, sas2}, so here  $\alpha_2(N,\beta)$ is a measure of clustering of the system. On the other hand the value of $\alpha_2(N,\beta)$ decreases by the temperature, hence value of internal energy is proportional to the temperature as illustrated by Fig. \ref{figu}.\\
We can see that the value of the internal energy is negative for the low temperature (see Fig. \ref{figu} (a)), however at high temperature we can see a transition to the positive value which is illustrated by Fig. \ref{figu} (b). It may be sign of a phase transition which should be examined by the specific heat analysis later. At very high temperature, the internal energy yields to a positive constant. In the Fig. \ref{figu} (c) we can see behavior of the low mass limit.

Now, it is possible to calculate the entropy of this system of galaxies from the internal energy of this system. Thus, we can write the entropy as
\begin{eqnarray}
\label{eq5.1}
{\cal S}=\left(\frac{3}{4}-\frac{97}{5\alpha_2(N,\beta)}\right)\left(\frac{8\pi^2}{3}\alpha_1(N,\beta)\alpha_2(N,\beta)\right)^{-\frac{1}{3}}
\end{eqnarray}
In order to have the positive entropy, the following condition is necessary,
\begin{equation}
\alpha_2(N,\beta)\geq\frac{388}{15}.
\end{equation}
Hence we can fix $\alpha_2(N,\beta)=26$ to obtain positive entropy and write,
\begin{eqnarray}
\label{eq5.2}
{\cal S}=\frac{1}{260}\left(\frac{208\pi^2}{3}\alpha_1(N,\beta)\right)^{-\frac{1}{3}}
\end{eqnarray}

\begin{figure}[h!]
 \begin{center}$
 \begin{array}{cccc}
\includegraphics[width=60 mm]{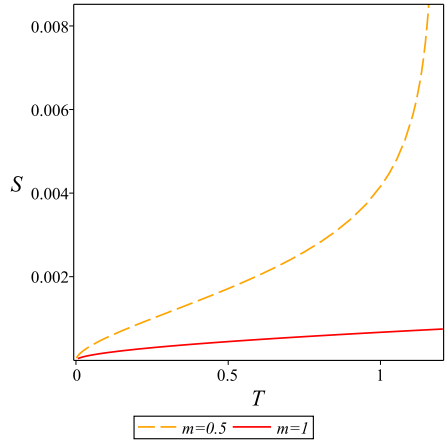}
 \end{array}$
 \end{center}
\caption{Typical behavior of the entropy in terms of temperature for $G=1$ and $N=50$.}
 \label{figs}
\end{figure}

\begin{figure}[h!]
 \begin{center}$
 \begin{array}{cccc}
\includegraphics[width=60 mm]{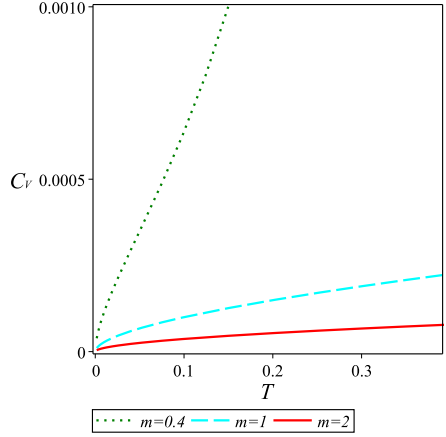}
 \end{array}$
 \end{center}
\caption{Typical behavior of the specific heat in terms of temperature for $G=1$ and $N=50$.}
 \label{figC}
\end{figure}

In the Fig. \ref{figs} we can see behavior of the entropy, which is increasing function of the temperature, while it is decreased by mass.\\
Then, we can study specific heat by using the following relation,
\begin{equation}
C_V=T\left(\frac{dS}{dT}\right)_{V}.
\end{equation}
We can see from the Fig. \ref{figC} that is increasing function of the temperature. We find that there is no negative specific heat and asymptotic behavior hence there is no phase transition and critical points. We show that the model is stable at all temperature. Heavier situations yields to the constant specific heat at higher temperatures.

Similarly, the  free Helmholtz energy can be expressed as
\begin{eqnarray}
\label{eq5.4}
{\cal F}&=&<{\cal U}>-T{\cal S}
 \\ \nonumber &=& -T\left[\frac{21}{4}+\frac{97}{5\alpha_2(N,\beta)}-\left(\frac{97}{5\alpha_2(N,\beta)}-\frac{3}{4}\right)\left(\frac{8\pi^2}{3}\alpha_1(N,\beta)\alpha_2(N,\beta)\right)^{-\frac{1}{3}}\right]
\end{eqnarray}
Finally, we can write the  pressure $P$ of this system as
\begin{eqnarray}
\label{eq5.8}
P&=& \frac {N} {V}\frac {2<{\cal U}>} {3N-2}\nonumber\\
&=&-\frac{9N} {2V\beta(3N-2)}\left(1+\frac{388}{45\alpha_2(N,\beta)}\right)
\end{eqnarray}
It is again known that the pressure of a system of galaxies  reduces as the galaxy cluster  \cite{sas, sas2}. Here we observe that as the value of
$\alpha_2(N,\beta)$ increases the pressure reduces. This again demonstrates that $\alpha_2(N,\beta)$  measure the clustering in a system of  galaxies, and can be used to analyze the large scale structure of our universe.

\setcounter{equation}{0}

\section{Correlation Function}
As the galaxies interact with each other, it is important to obtain the correlation function between various galaxies. So, we can now obtain the correlation function between various galaxies as
\begin{equation}
\label{eq6.1}
\zeta=\int\xi dV= -\frac{NT}{V^2}\left(\frac{\partial V}{\partial P}\right)_T-1.
\end{equation}
Now we can write this as
\begin{equation}
\label{eq6.2}
\zeta=\int\xi dV= -\frac{2(3N-2)}{k}\left(1+\frac{388}{45\alpha_2(N,\beta)}\right)_T-1
\end{equation}
So, for ($k =1$), we can express the correlation function for a system of galaxies as
\begin{equation}
\label{eq6.3}
\zeta=\int\xi dV= \left[(3-6N)+(4-6N)\frac{388}{45\alpha_2(N,\beta)}\right].
\end{equation}

Due to this correlation between galaxies, individual galaxies can merge with each other, changing the number of galaxies in the system. Furthermore,  as the universe expands, the number of galaxies in the observable universe, or any specific volume $V$ change. This can can be measured by using the  grand canonical partition function
\begin{equation}
\label{eq7.1}
{\cal Z}_{G\nu}=\int\limits_{-\infty}^{\infty}d^\nu x\int\limits_{-\infty}^{\infty}d^\nu p
\left[1+(q-1)\beta\left(\frac {N(N-1)Gm^2} {2r}-\frac {Np^2} {2m}+\mu N\right)\right]_+^{\frac {1} {q-1}},
\end{equation}
where $\mu$ is the chemical potential for this system.
This can in turn be expressed as
\begin{eqnarray}Z_{G\nu}&=& \frac {1} {2}\left[\frac {2\pi^{\frac {\nu} {2}}} {\Gamma\left(\frac {\nu} {2}\right)}\right]^2
\left[\frac {m} {2\beta(q-1)N}\right]^{\frac {\nu} {2}}
[(q-1)\beta N(N-1)Gm^2]^\nu
\\ \nonumber &&  \times
(1+\beta\mu N)^{\frac {\nu} {2}+\frac {1} {q-1}}B\left(1+\frac {1} {q-1},\frac {\nu} {2}\right)
B\left(\frac {\nu} {2}-\frac {1} {q-1},-\nu\right)
\\ \nonumber &=&\frac {2^{1-\frac {\nu} {2}}\pi^\nu} {\Gamma\left(\frac {\nu} {2}\right)^2} 3^{-\frac {\nu} {2}}
(mN\beta)^{\frac {\nu} {2}}
[(N-1)Gm^2]^\nu
\\ \nonumber &&  \times  (1+\beta\mu N)^{\frac {\nu} {2}+3}B\left(4,\frac {\nu} {2}\right)
B\left(\frac {\nu} {2}-3,-\nu\right)
\end{eqnarray}
This partition function also diverges. However, as it is possible to write the grand canonical partition function as
\begin{equation}
\label{eq7.4}
Z_{G\nu}=Z_{\nu}(1+\beta\mu N)^{\frac {\nu} {2}+3}
\end{equation}
So, we can again use GDR to evaluate the  grand canonical partition function  as
\begin{eqnarray}
\label{eq7.5}
Z_{G}&=& (1+\beta\mu N)^{\frac {9} {2}}\left[-f(3)C-f^{\prime}(3)-\frac{f(3)}{2}\ln(1+\beta \mu N)\right]
\\ \nonumber &=&  (1+\beta\mu N)^{\frac {9} {2}}\left[{\cal Z}+\frac {8\pi^2} {\sqrt{3}}\left(\frac {mN\beta} {2}\right)^{\frac {3} {2}}
[m^2G(N-1)]^3\ln(1+\beta\mu N)\right]
\end{eqnarray}
This is a finite value for the  grand canonical partition function for a system of galaxies. Here again the divergences in the grand canonical partition function  have been removed using GDR. It would be possible to use this grand canonical partition function to obtain various thermodynamic quantities for a system of galaxies, where the number of galaxies change due to high correlation between them.

\section{Conclusion}

In this paper, we  have analyzed the clustering of galaxies using the techniques of statistical mechanics. As the size of galaxies is small compared to the distance between them, individual galaxies can be approximated as point particles. These galaxies interact through a gravitational potential, and it is known  that for such a system  a violation of the   extensive property occurs, and use of Boltzmann-Gibbs statistical mechanics to study such a system becomes a constraint. However, as the  Boltzmann-Gibbs statistical mechanics has been generalized to Tsallis statistical mechanics, and this property is not required in Tsallis statistical mechanics, thus paved the way to  use Tsallis statistical mechanics for analyzing such a system.
The divergences in the Tsallis gravitational partition function are eliminated by using the powerful techniques of GDR which is the important part of this paper. In fact the important conclusion to be drawn here is that the Tsallis statistics supplemented with the generalization of  dimensional regularization makes the work mathematically rich and a strong treatment to deal with clustering of galaxies or the analysis of gravitating systems in the expanding universe. We especially used GDR to obtain finite gravitational partition function devoid of any divergences for analyzing thermodynamics of the gravitating system, and thus to study its relation to clustering of galaxies. We also analyzed the correlation function between galaxies for this system. The correlation function is believed to contain the complete information of clustering. The grand partition function obtained is also treated with GDR for eliminating the divergences. Thus we conclude that the thermodynamics obtained in this paper comes from the divergence free partition functions and the interesting results are depicted from the plots. The typical behavior in different temperature ranges is worth mentioning.

\end{document}